\documentclass[times,10pt,a4paper]{article}

\usepackage[utf8]{inputenc}
\usepackage{graphicx}
\usepackage{nopageno}
\usepackage{textcomp}
\usepackage{pdfpages}
\usepackage{multirow}
\usepackage{listings}
\usepackage{float}
\usepackage[caption=false]{subfig}
\usepackage{graphics}
\usepackage{array}
\usepackage{color}
\usepackage{midfloat}
\usepackage[cmex10]{amsmath}
\usepackage{moreverb}
\usepackage{cite}
\usepackage{fixltx2e}
\usepackage{url}
\usepackage{fancyhdr}
\usepackage{rotating}
\usepackage{algorithmic}
\usepackage{algorithm}
\usepackage{paralist}
\usepackage{amssymb}
\usepackage[affil-it]{authblk}
\usepackage{fullpage}

\makeatletter
\renewcommand*{\@fnsymbol}[1]{\ensuremath{\ifcase#1\or \dagger\or \ddagger\or
   \mathsection\or \mathparagraph\or \|\or **\or \dagger\dagger
   \or \ddagger\ddagger \else\@ctrerr\fi}}

\makeatother

\begin{document}
\date{}
	
\title{Robustness surfaces of complex networks}    

\author{Marc Manzano$^{1,}$\thanks{mmanzano@eia.udg.edu} $^{,}$\thanks{This work was done while visiting the EPICENTER research group at Kansas State University, USA.} , Faryad Sahneh$^2$, Caterina Scoglio$^2$,\\ Eusebi Calle$^1$, Jose Luis Marzo$^{1,2}$}     
\affil{$^1$Department of Architecture and Computers Technology, University of Girona, Spain\\$^2$Department of Electrical and Computer Engineering, Kansas State University, USA}

\maketitle

\begin{abstract}
{\bf Despite the robustness of complex networks has been extensively studied in the last decade, there still lacks a unifying framework able to embrace all the proposed metrics. In the literature there are two open issues related to this gap: (a) how to dimension several metrics to allow their summation and (b) how to weight each of the metrics. In this work we propose a solution for the two aforementioned problems by defining the $R^*$-value and introducing the concept of \emph{robustness surface} ($\Omega$). The rationale of our proposal is to make use of Principal Component Analysis (PCA). We firstly adjust to 1 the initial robustness of a network. Secondly, we find the most informative robustness metric under a specific failure scenario. Then, we repeat the process for several percentage of failures and different realizations of the failure process. Lastly, we join these values to form the robustness surface, which allows the visual assessment of network robustness variability. Results show that a network presents different robustness surfaces (i.e., dissimilar shapes) depending on the failure scenario and the set of metrics. In addition, the robustness surface allows the robustness of different networks to be compared.}
\end{abstract}

%\keywords{Robustness; Complex Networks; Uniformed metric}
\section*{}

%Critical infrastructures (e.g., power grids or telecommunication networks), biological networks (e.g., protein interaction networks), social networks (e.g., scientific authors collaboration networks), economic networks (e.g., trans-national credit and investment networks), 

The study of complex networks has attracted significant attention in the past decade. Critical infrastructures such as power grids, telecommunication networks or transportation networks, among others, are complex networks which are omnipresent and play a pivotal role in ensuring the smooth functioning of modern day living. These networks have to constantly deal with failures of their components, hence, any disruption of the service provided might have a considerable impact upon sizable proportions of the world’s inhabitants. Thus, understanding not only the structure, but also the dynamics of such networks is of paramount importance. 

Failures can be classified as being either random (i.e., accidental) or intentional (also referred to as targeted or deliberated) \cite{albert2000error,Manzano2013endurance}. Accidental failures occur as a result of random actions on network elements (e.g., human-made errors or natural disasters). In contrast, in intentional attacks components are chosen according to some criterion in order to maximize the impact of the failures (e.g., a Denial-of-Service (DoS) attack). We define a \emph{failure scenario} as the pair given by a specific type of failure (e.g., node or link) and a given attack strategy (e.g., random or intentional). %, and a percentage of elements failed/attacked (e.g., 10\%).

For network engineers and operators it is crucial to quantify the tolerance of a network to a given failure scenario. Robustness is defined as the ability of a network to maintain its total throughput under node or link removal \cite{Sydney2010elasticity,manzano2011quantitative}. 

Robustness metrics have been evolving since the advent of network science. Initially, several works studied the robustness of complex networks by considering a single graph metric: efficiency \cite{Lattora2001efficient}, average shortest-path length \cite{smallWorld,shannon2004spread}, diameter \cite{albert1999internet}, clustering coefficient \cite{smallWorld,Bollobas03cc}, node and link connectivity \cite{dekker2004network}, heterogeneity \cite{dong2007understanding}, two-terminal reliability \cite{A2TR}, assortativity \cite{Mahadevan:2006:IAT:1111322.1111328}, betweenness centrality \cite{freeman1977bc}, among others. Later on, new metrics were proposed in order to capture advanced characteristics (i.e., by means of spectral graph theory): symmetry ratio \cite{dekker2005symmetry}, algebraic connectivity \cite{JamakovicM08} or spectral radius \cite{vanmieghem2009virusspread}. Furthermore, other works presented more contemporary metrics which were based on classical graph features. For instance, the authors of \cite{Rohrer2011RNDM} studied the robustness in terms of flow diversity, a metric based on the shortest-path length. More recently, generic procedures to capture the robustness of a network for the whole spectrum of possible failures have been presented. Metrics such as elasticity \cite{Sydney2010elasticity} or endurance \cite{Manzano2013endurance} quantify the robustness of a network according to a single throughput parameter. Trajanovski et al., have proposed a framework to evaluate the robustness of complex networks, which is based on the generic metric $R$-value \cite{Trajanovski26032013}. From now on, we will use the conventions defined in Table~\ref{tab:var_def}. According to \cite{pvannieghemframework}, the $R$-value is denoted by:
\begin{equation} \label{pieteq}
	R = \sum_{k=1}^{n}s_k t_k
\end{equation}  
where $s$ and $t$ are $n \times 1$ weight and graph metric vectors, respectively, and $n$ is the number of robustness metrics. Thus, the $R$-value includes several graph metrics characterizing network robustness. However, there are two open issues related to the normalization of the $t$ metrics:
\begin{enumerate}
	\item How to unify the dimensionality of each robustness metric of vector $t$ in order to legitimate their summation.
	\item How to define the weight of each metric to optimally extract the most significant information.
\end{enumerate}

In this work we propose a solution for the two aforementioned problems by defining the $R^*$-value and introducing the concept of \emph{robustness surface} ($\Omega$). The former extracts the most informative robustness metric for a failure scenario, while the latter allows network robustness variations of different networks to be visually assessed, regardless of the failure scenario.

\section*{Results}

\textbf{$R^*$-value.} The rationale of our proposal is to make use of Principal Component Analysis (PCA) (see \emph{Methods}). Given a set of robustness metrics $t$, we first define the initial robustness as follows:
\begin{equation} \label{rStar}
	R^{*}_{init} = \sum_{k=1}^n \hat{v_k}t^0_k = 1
\end{equation}
where $t^0$ is the set of metrics when no failures occur, and $\hat{v}$ is a normalized eigenvector or Principal Component (PC). We obtain $\hat{v}$ from the procedure that computes the robustness surface (see following subsection and Eq.~\ref{normV}). The fact that $\hat{v}$ is normalized makes $R^{*}_{init}$ equal to 1. Additionally, $R^*$ can be computed when $p$\% of elements fail as denoted next:
\begin{equation} \label{rStarP}
	R^{*}_{p} = \sum_{k=1}^n \hat{v_k}t^p_k
\end{equation}
where $t^p_k$ is the set of metrics computed when $p$\% of failures occur. $R^{*}_{p}$ takes values in the interval [0,+$\infty$).

The difference between $R^*$ and $R$ (Eq.~\ref{pieteq}) is that in our proposal the principal component $\hat{v}$ gives dimension and non-arbitrary weights to each of the metrics. In addition, besides finding the most informative robustness metric, we adjust the initial robustness to 1, thus simplifying the comparison of network robustness variations when failures occur.

\textbf{Robustness surface ($\Omega$).} The robustness surface allows the network performance variability for a given failure scenario to be visually assessed. 

In fact, $\Omega$ is a matrix where the rows are the percentage of failures ($P$) and the columns are the distinct failure configurations ($m$). The list of percentage of failures $P$ (e.g., $P=\{1\%,2\%..100\%\}$) denotes the range of failures for which the robustness is evaluated. A \emph{failure configuration} represents a realization of the failure process. The different failure configurations $m$ depict the different subsets of elements that fail for a given percentage of failures, with each subset being distinct from one another. The robustness value in $\Omega[p][i]$, where $p \in \{1\%..|P|\%\}$ and $i \in \{1..m\}$, is given by $R^*_p$ (Eq.~\ref{rStarP}). 

To obtain the robustness surface of a network given a failure scenario (e.g., node and random), we define the following procedure:
\begin{enumerate}
\item Let $A_p$ be an $m \times n$ matrix where $p \in \{1\%..|P|\%\}$ is the percentage of failure. The goal is to transform $A_p$ into a smaller data set, i.e., a vector $\omega_p$ of size $m$, while preserving the most significant information. Therefore, we define $\omega_p$ as a vector of size $m \times 1$. $\omega_p$ contains the set of $m$ values $R^*_p$ computed when $p$\% of elements fail. 

\item To do so, we first compute the covariance matrix $C_p$ of each matrix $A_p$. Then, we average the $|P|$ covariance matrices to obtain a unique matrix $\bar{C}$. This allows us to obtain a PC independent of $p$. 

\item We calculate the eigenvectors $V$ and the eigenvalues $D$ of $\bar{C}$. At this point, the $l$ most relevant eigenvectors of $V$ are taken as the principal components for each matrix $A_p$ (see \emph{Methods} for further details). Hereafter we assume that $l=1$, i.e., $v$ is the eigenvector PC.

\item Then, we obtain $\hat{v}$ by normalizing $v$:

	\begin{equation} \label{normV}
		\hat{v}_j = \frac{v_j}{\sum_{k=1}^{n} t^0_k v_k} \quad \quad j \in \{1..n\}
	\end{equation}

\item By multiplying the principal component $\hat{v}$ by each row of $A_p$ we obtain a vector $\omega_p$ of size $m$. Each value of $\omega_p$ is, indeed, $R^*_p$. Next, by iterating this procedure for all matrices $A_p$, we obtain a set of $|P|$ vectors $\omega_p$. Finally, we define ${\omega}'_p$ as a vector $\omega_p$ sorted in decreasing order. Consequently, the robustness surface is given by the following expression $\Omega = \{{\omega}'_{1\%},...,{\omega}'_{|P|\%}\}$.
\end{enumerate}

Although different failure scenarios (e.g, link random and link by betweenness centrality) provide different $\hat{v}$, each of them satisfies Eq.~\ref{rStar} because $\hat{v}$ is normalized (as shown in Eq.~\ref{normV}).

\textbf{Case study.} Here, we illustrate the suitability of our proposal for evaluating the robustness when considering several metrics. To do so, we study two real critical infrastructures: the Spanish railway network (\emph{sprailway}) \cite{Roanes-Lozano2009SpanishRailway} and the European power grid network (\emph{europg}) \cite{bialek2013}. 

We consider incremental and irreversible random and targeted attacks (e.g., betweenness centrality (BC) or node degree). Link and node failures are considered for the \emph{sprailway} and \emph{europg}, respectively, to show that the robustness surface allows us to compare network robustness independently from the failure scenario. Link failures are caused randomly and by link BC, whereas node failures are caused randomly, by node degree, the clustering coefficient and node BC. In both cases, $|P|$ is set to 70, i.e., from 1\% to 70\% of failures. The presented results are obtained for 500 and 100 runs ($m$) for random and targeted attacks, respectively. For each of the runs, a different realization of the failure process is considered, i.e., a distinct subset of elements that fail according to the failure scenario.
%These values of $m$ are examples, and other studies might require different values.

We consider the following metrics: the largest connected component (LCC), the degree of fragmentation as a function of the number of connected components (only applicable to link failures), the average nodal degree, the two-terminal reliability, the average clustering coefficient, the average shortest-path length, the diameter, the average node BC, the average link BC and the algebraic connectivity. Therefore, link failures have $n=10$ while node failures have $n=9$.

Table~\ref{tab:features} presents the main characteristics of the two networks considered. Although both networks have a different number of nodes and links, they show a similar average node degree $\langle k \rangle$, which is between 2 and 3. However, \emph{europg} has a higher node maximum degree, what means that such a network is more vulnerable to targeted attacks. The average shortest-path length $\langle l \rangle$ depicts that \emph{europg} is about two times wider than \emph{sprailway}. Finally, both networks have a negative value of assortativity ($r$), which means that nodes of dissimilar degrees are connected to each other. 

\begin{figure}
 \centering

\subfloat[\emph{sprailway} random]{
 \includegraphics[scale=0.4]{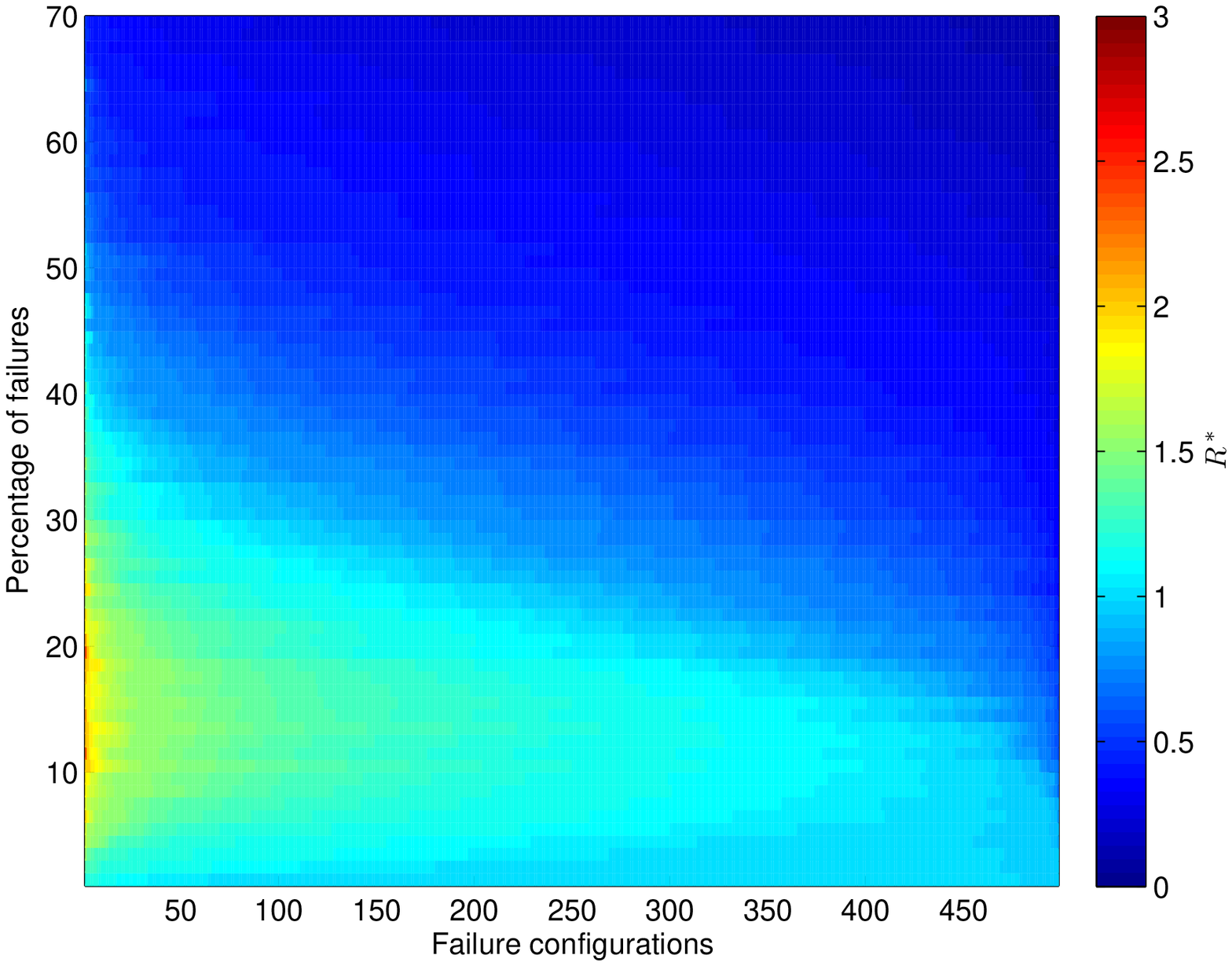}
  \label{fig:sprailway_rnd}
}
  \subfloat[\emph{sprailway} link betweenness centrality]{
   \includegraphics[scale=0.4]{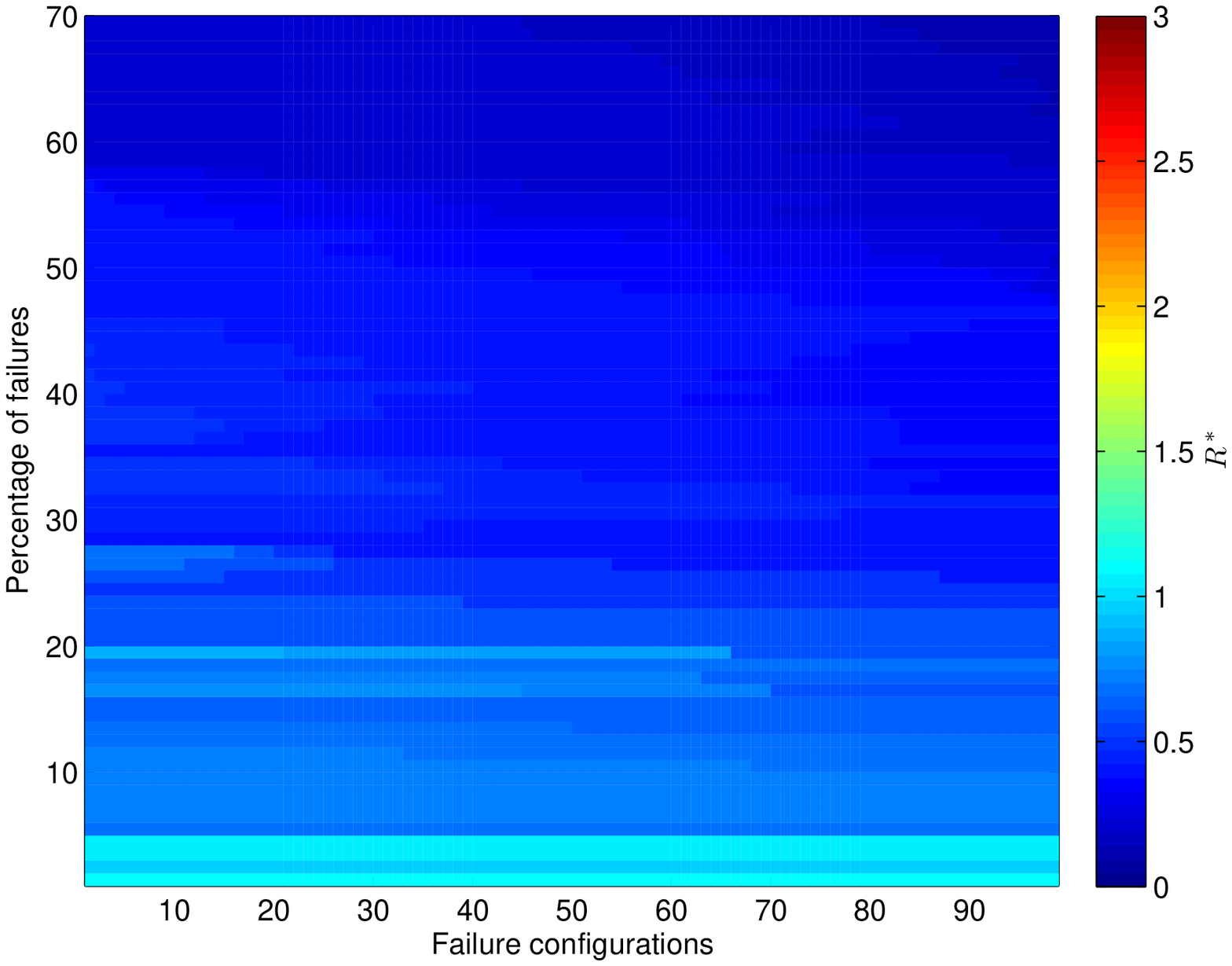}
    \label{fig:sprailway_bc}
  }

  \caption{Robustness surface $\Omega$ of \emph{sprailway} when causing links to fail randomly and by link BC.}
  \label{fig:sprailway_res}
\end{figure}

\begin{figure}
 \centering

\subfloat[\emph{europg} random]{
 \includegraphics[scale=0.4]{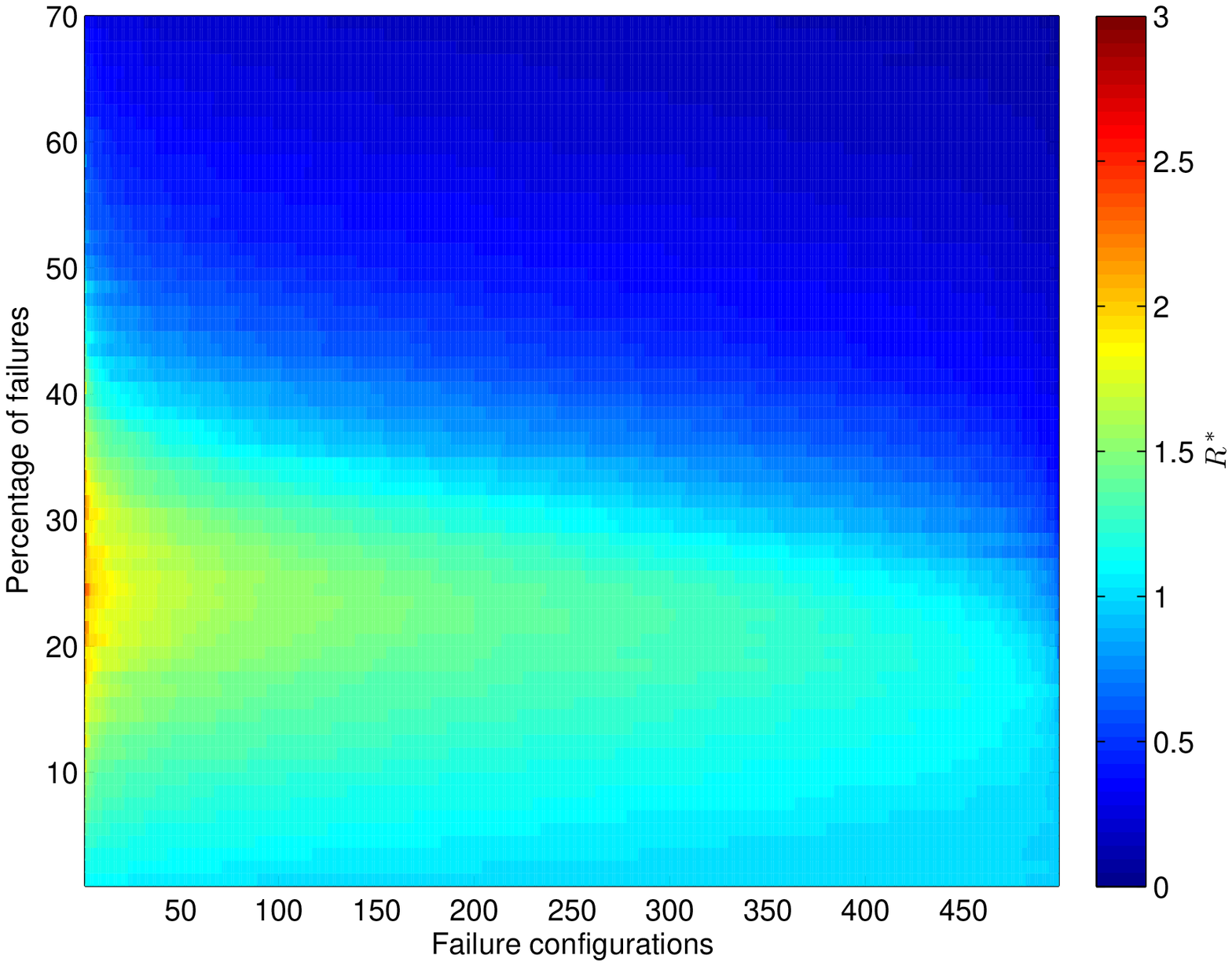}
  \label{fig:europg_rnd}
}
  \subfloat[\emph{europg} node betweenness centrality]{
   \includegraphics[scale=0.4]{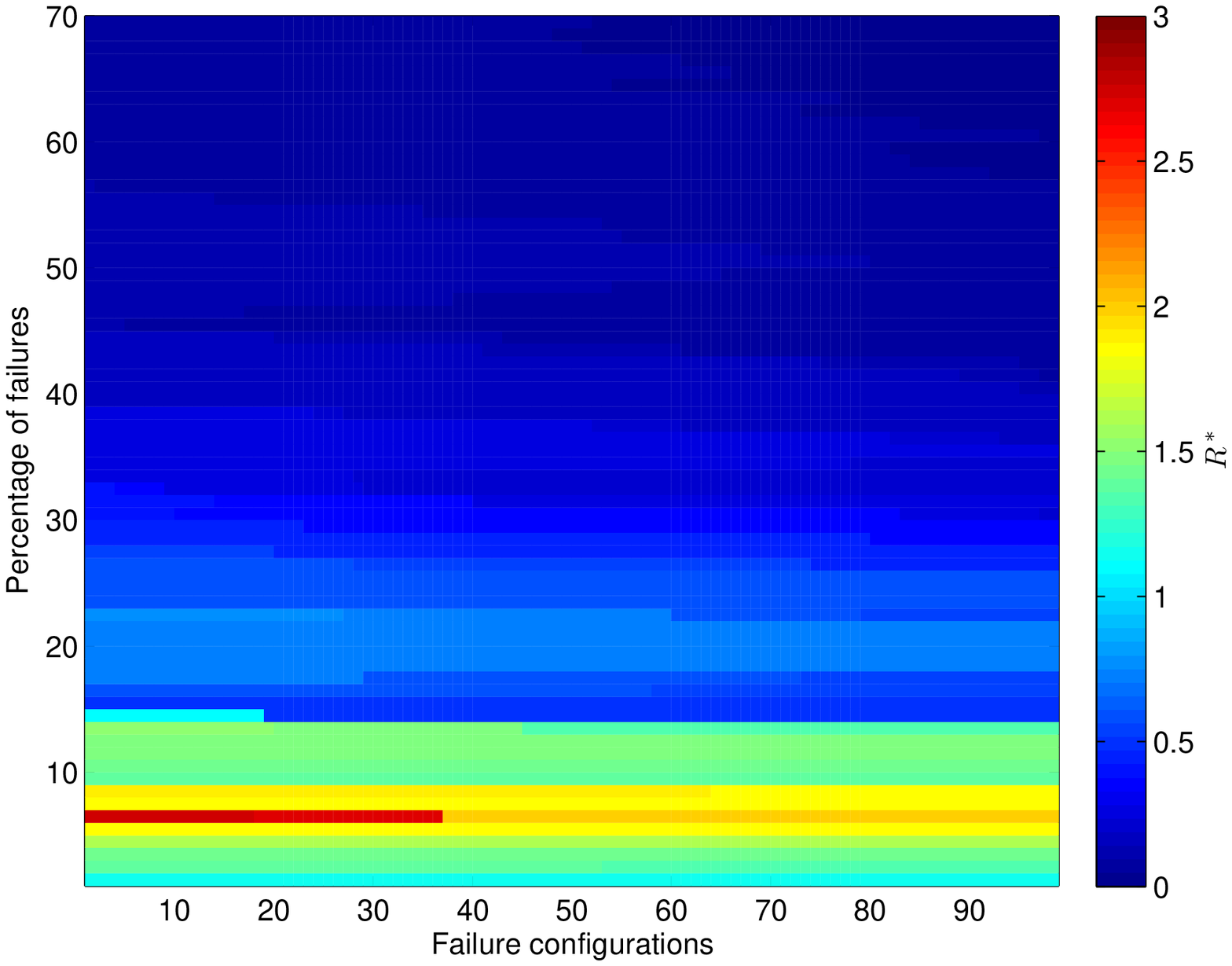}
    \label{fig:europg_bc}
  }
\\
\subfloat[\emph{europg} node degree]{
 \includegraphics[scale=0.4]{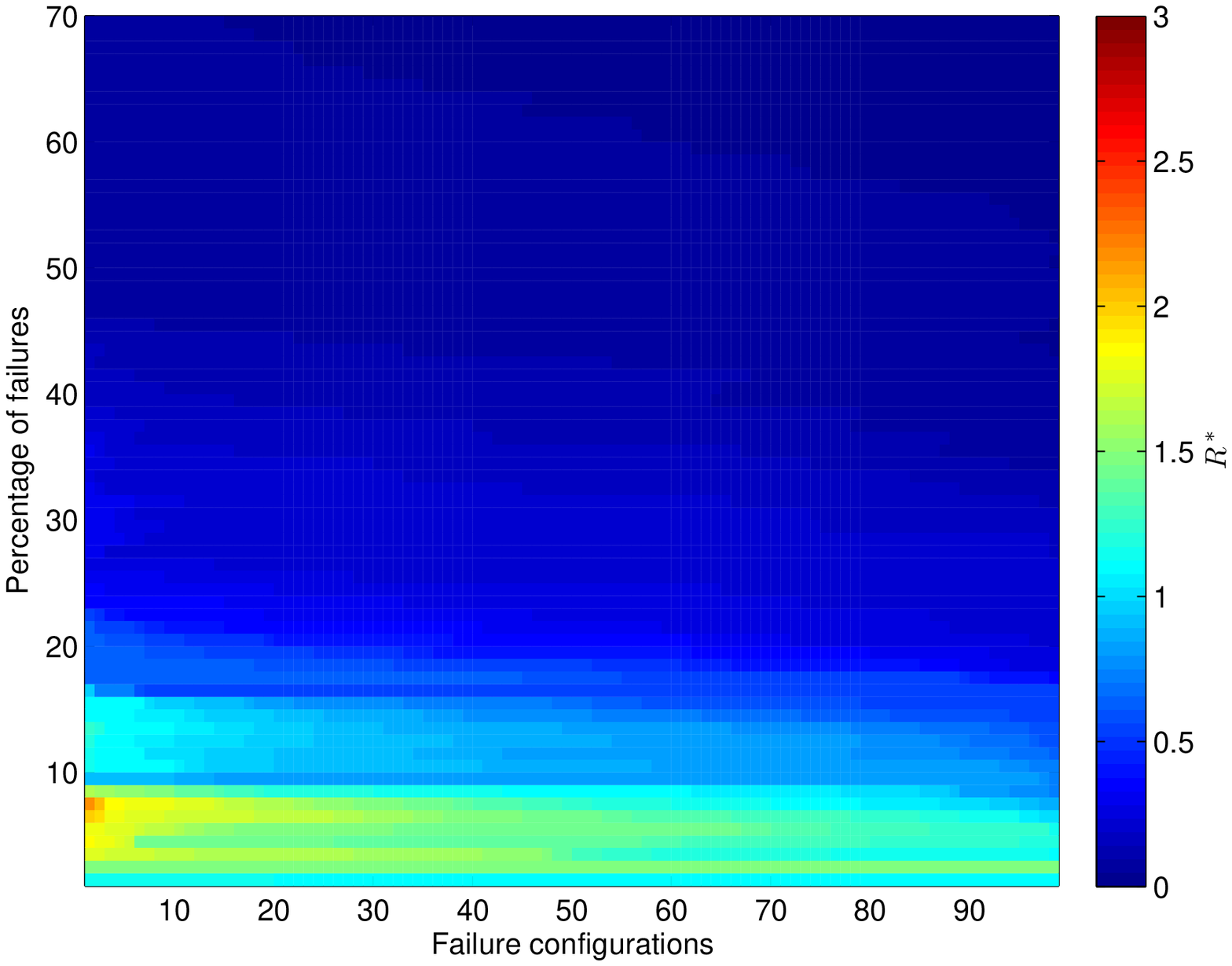}
  \label{fig:europg_deg}
}
  \subfloat[\emph{europg} node clustering coefficient]{
   \includegraphics[scale=0.4]{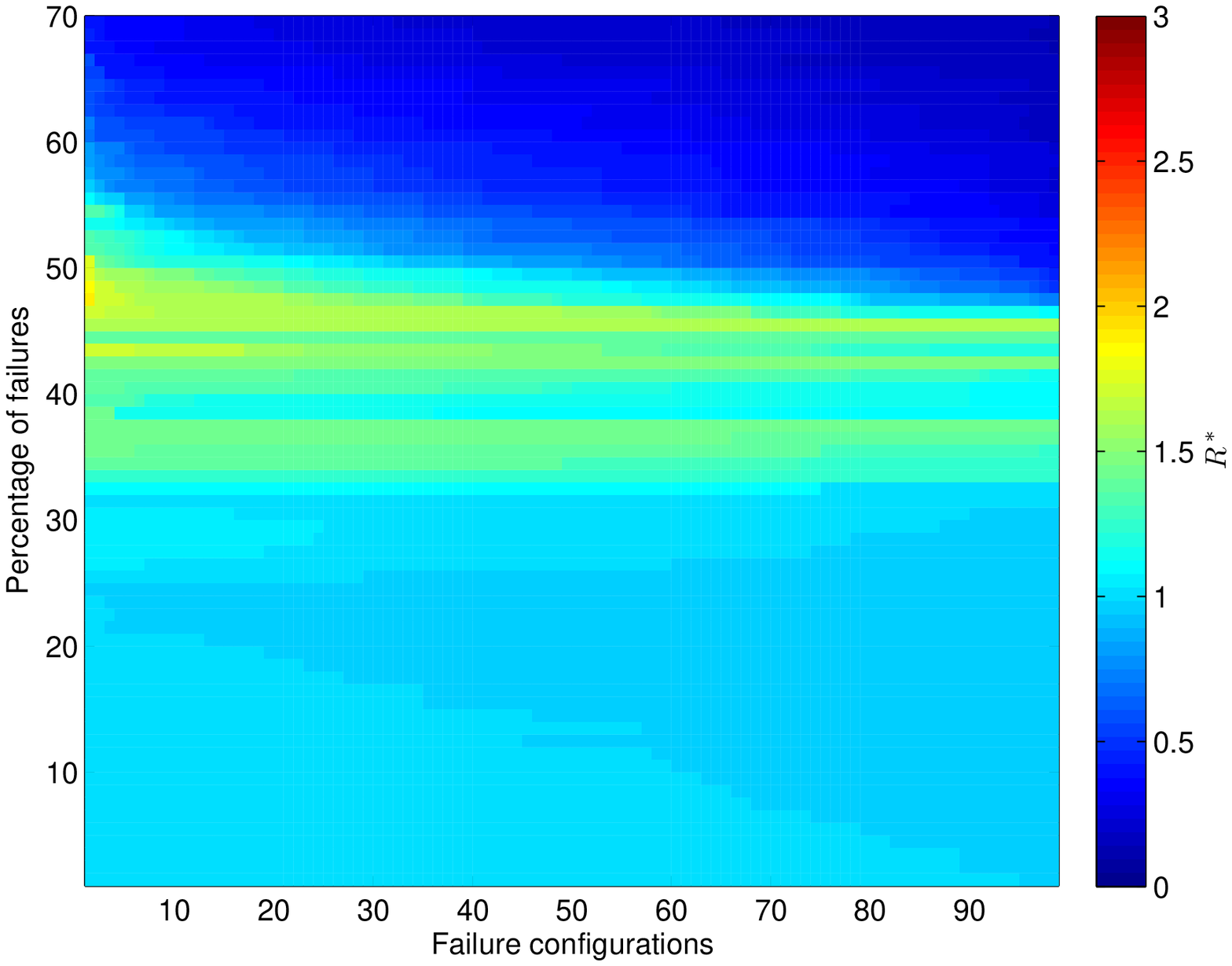}
    \label{fig:europg_cc}
  }
  \caption{Robustness surface $\Omega$ of \emph{europg} when causing nodes to fail randomly, by node BC, by node degree and by the clustering coefficient.}
  \label{fig:europg_res}
\end{figure}

\textbf{Numerical results.} The results of our work are presented in Figures \ref{fig:sprailway_res} and \ref{fig:europg_res}. The x-axes show the different failure configurations for which the $n$ metrics have been computed. The y-axes depict the range of percentage of failures (from 1\% to 70\%). At each coordinate (x,y), i.e., for each percentage of failures and for each subset of elements that fail, the $R^*_p$-value is shown. In each figure the range of colors expresses variability, with dark blue and dark red being the two extremes of each failure scenario intervals. Since $R^*_{init}=1$, i.e., the initial robustness is set to 1 regardless of the set of $n$ chosen metrics, our results allow a visual assessment of the robustness variation with respect to the initial conditions. The further the value of $R^*_p$ is with respect to $R^*_{init}$, the lower the performance of the network is. When $R^*_p$ is close to 0, the performance is considered to be totally deteriorated. Moreover, it is possible to observe $R^*_p=1$ when $p\ge1\%$, and the LCC of the network has similar properties to the initial network (without failures). 

Figure~\ref{fig:sprailway_res} presents the robustness surface $\Omega$ of \emph{sprailway} in the case of random (Fig.\ref{fig:sprailway_rnd}) and link BC failures (Fig.\ref{fig:sprailway_bc}). Interestingly, the random case provides a smooth surface, while the targeted case presents abrupt slopes. The latter is worth noting, because the presence of abrupt slopes in the robustness surface means that there are network elements (in this case, links) that could be protected in order to improve the overall network robustness.

In the case of \emph{europg}, Fig.~\ref{fig:europg_res} depicts four robustness surfaces under different node failure scenarios. Similar to \emph{sprailway}, the random surface depicts a regular behavior. In addition, the targeted-based cases depict rough surfaces. While Figs.~\ref{fig:europg_bc} and \ref{fig:europg_deg} depict that \emph{europg} is not robust under node degree or node BC failure scenarios, Fig.~\ref{fig:europg_cc} shows that the network keeps the initial robustness until more than 30\% of the nodes have failed. This implies that \emph{europg} is significantly more robust under failures by the clustering coefficient than by other targeted strategies.

For some failure configurations, it is worth noting that $R^*_p$ might increase at some percentage of failures with respect to $R^*_{init}$, as observed in 10\% or 20\% of failures in Fig.~\ref{fig:sprailway_rnd} and in 20\% or 30\% in Fig.~\ref{fig:europg_rnd}, as well as in the targeted-based surfaces. This result should not be misleading because it totally depends on the set of metrics that are being considered for the study. For instance, while some metrics might decrease as the percentage of failures increases, others might alternate increments and decrements because they depend on the number and size of largest connected components (i.e., average shortest-path length, diameter, algebraic connectivity, etc.). Therefore, the suitability of the robustness surface remains intact, because the variability of the robustness can be assessed in any case.

\begin{figure}
 \centering
 
 \subfloat[Mean: $\bar{R}^{*}$]{
 \includegraphics[scale=0.43]{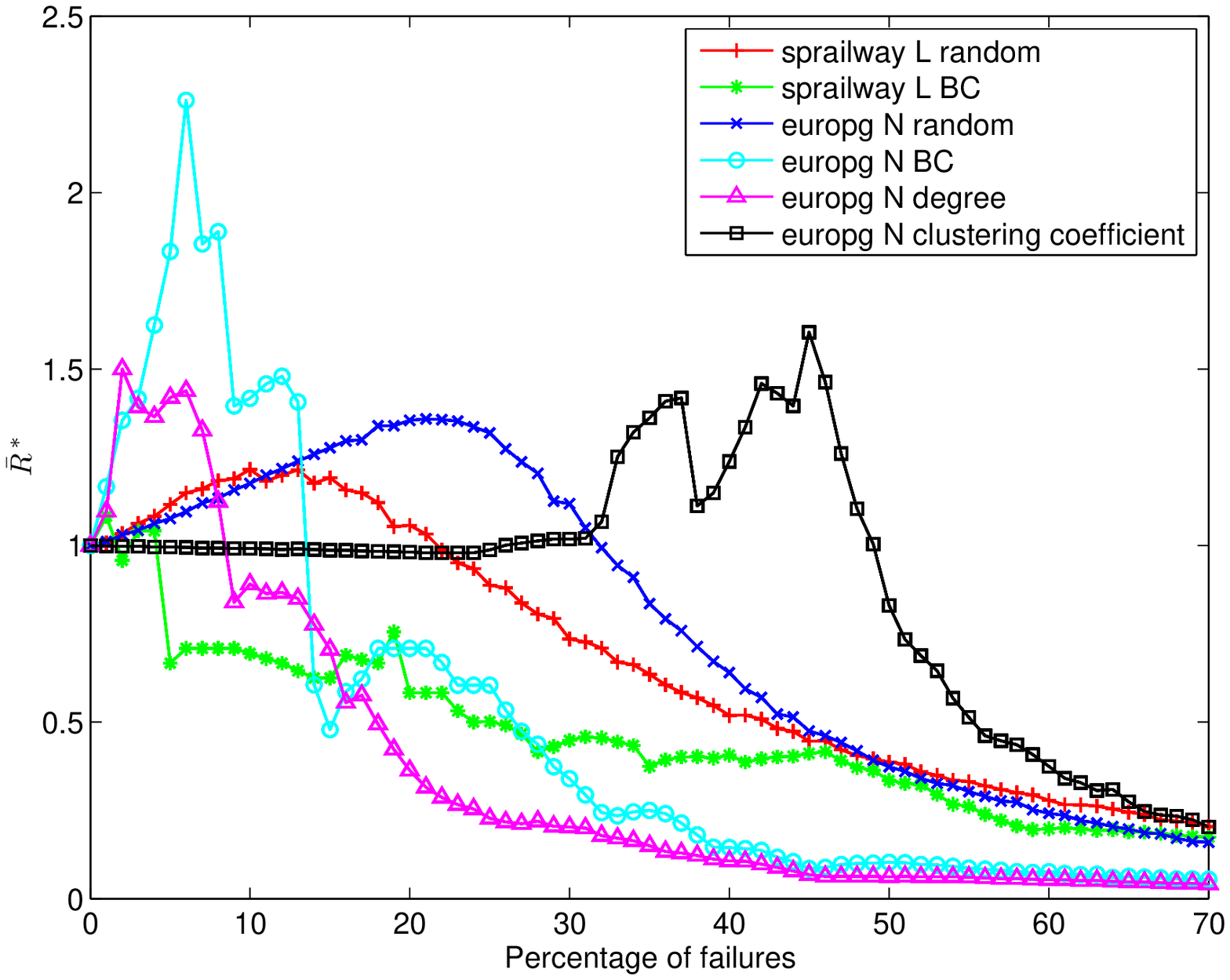}
   \label{fig:summary_avg}
 }
 \subfloat[Variance: $R^{*}$]{
  \includegraphics[scale=0.43]{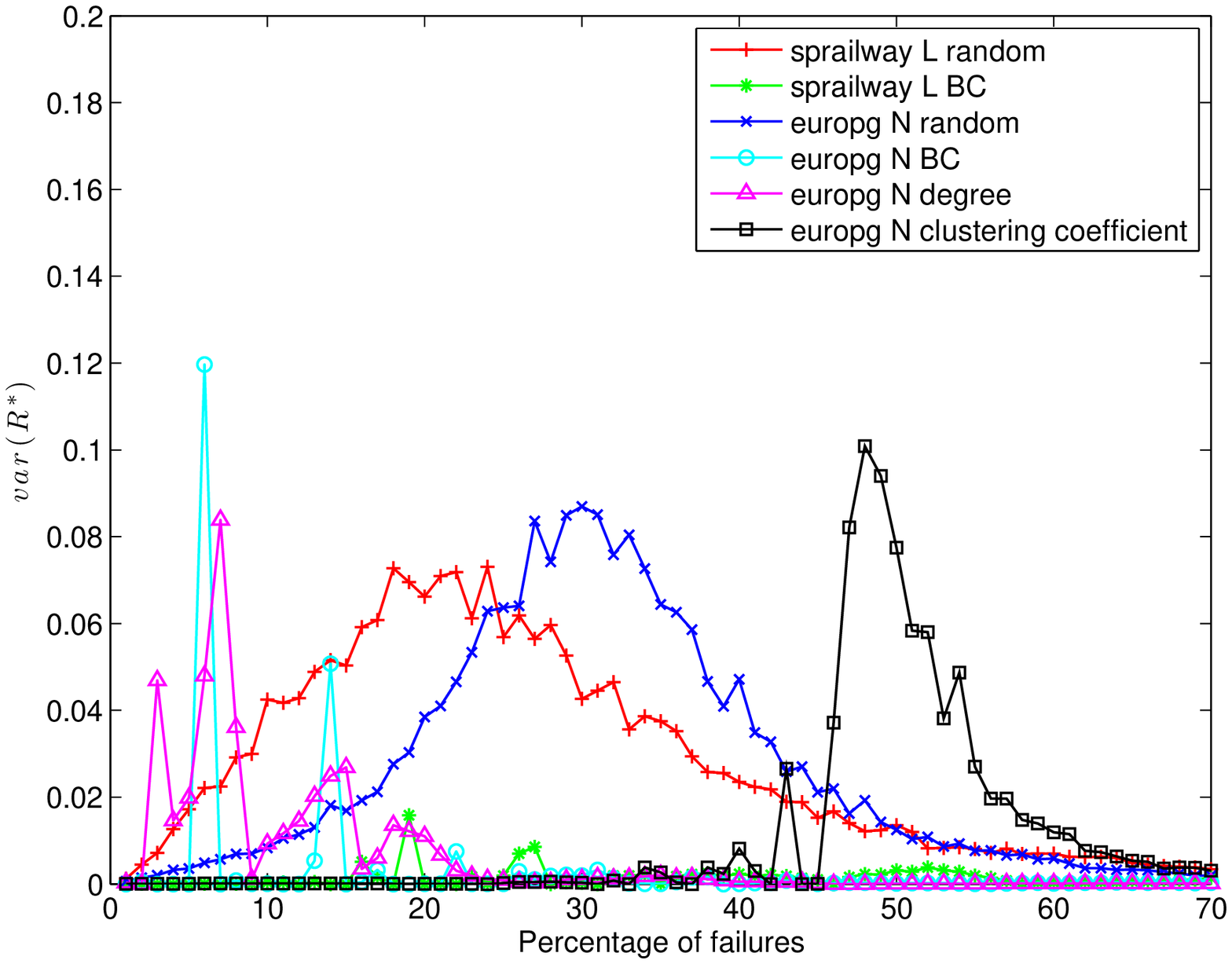}
   \label{fig:summary_var}
 }
 \caption{Robustness summary of \emph{sprailway} and \emph{europg} under the different failure scenarios. As to the legend, L refers to link failures, whereas N refers to nodes.}
  \label{fig:summary}
\end{figure}

Finally, to compare the robustness surfaces of both networks, and considering the different failure scenarios, we average the values of each ${\omega}'_p$ of $\Omega$. Thus, for each network and failure scenario, we obtain $|P|$ $\bar{R^*_p}$-values. Fig.~\ref{fig:summary} depicts a summary of the results. Fig.~\ref{fig:summary_avg} shows the curves of $\bar{R^*_p}$ of both networks from 1\% to 70\% of failures. To complement the results in Fig.~\ref{fig:summary_avg}, the variance is presented in Fig.~\ref{fig:summary_var}. For instance, it can be observed that both random failure scenarios show similar behaviors, although for \emph{europg} the top of the curve is around 24\% of failures. Therefore, our approach allows us to compare different networks, regardless of the failure scenarios. This comparison could be done numerically, for instance, by comparing the areas below the curves.

\section*{Discussion}

In this work we present the $R^*$-value and the concept of \emph{robustness surface} ($\Omega$). The rationale of our proposal is to make use of Principal Component Analysis (PCA). 

The $R^*$-value solves two open issues in the robustness of complex networks field. Our proposal extracts the most significant information from a set of robustness metrics. $R^*$ is the first generic metric able to characterize the robustness of complex networks with a single value, while taking into account several robustness metrics.

The robustness surface $\Omega$ provides a framework to visually assess the network robustness variability. Moreover, it allows for the comparison of the robustness between different networks under distinct failure scenarios. To the best of our knowledge, it is the first method of its kind to allow the visual evaluation of the network robustness for a specific failure scenario, while at the same time considering several robustness metrics.

Robustness surfaces are designed as a visual monitoring tool. First, our approach is applicable to real-time monitoring of a network through a single value, when it is otherwise implemented according to multiplicity of correlated metrics with possible inherent redundancy. Second, $\Omega$ can be a pivotal part of a network robustness refinement process:
\begin{itemize}
	\item Step 1: If the robustness surface presents abrupt slopes, then there are network elements (nodes or links) which are weaker than the rest, for a given failure scenario. These elements could be identified by means of traditional robustness metrics such as the betweenness centrality.
	\item Step 2: Enhance or protect the weak elements, for instance, by adding new links or applying immunization techniques.
	\item Step 3: Re-evaluate the robustness of a network and, instead of comparing a large number of robustness metrics, detect through visual inspection if the network robustness has been improved. 
\end{itemize}

We believe that the contributions presented in this work will lay a firm foundation for future research on the robustness of complex networks.

To conclude, the $R^*$-value shows that there is no single and universal robustness metric for a network. Instead, the robustness varies according to the failure scenario and the metrics that are used to quantify the performance of the network.

For future work, we plan to study the stability of the robustness surfaces with respect to network size scaling.

\section*{Methods}
\textbf{Principal Component Analysis (PCA).} PCA is a powerful tool to identify the most significant information in a data table representing observations described by several dependent variables, which can be inherently correlated. The goals of the PCA are to: (a) extract the main information of a data set and express it by means of new orthogonal variables called principal components; and (b) compress the size of the data set while preserving the most important information \cite{Abdi10principalcomponent}.

Let $A$ be a data set of $m$ observations of a vector-valued variable, i.e., $A \in R^{m \times n}$. We define $C$ as the covariance matrix of $A$, which is denoted by:
\begin{equation}
	C^{n \times n} = (c_{i,j},c_{i,j} = cov(col_{A_i},col_{A_j}))
\end{equation}
where $i,j \in \{1..n\}$, and $cov(col_{A_i},col_{A_j})$ is the covariance function evaluating column $i$ and column $j$.

PCA works with the spectrum of $C$. Let $v_i \in R^{n \times 1} \{i \in 1..n\}$ and $\lambda_i \in R$ be the eigenvectors and corresponding eigenvalues of the covariance matrix $C$, respectively. The matrix $V$ with all $v_i$ as columns represents the principal components, and provides an orthogonal transformation to the PC space. Furthermore, we denote $D$ as a matrix with the eigenvalues in its diagonal.

Let $\tilde{V}$ be $n \times l$ matrix, which only contains the top $l$ of the most important principal components (see \emph{Methods: Most relevant principal components of $A$} for further details). Therefore, we can obtain the transformed data $\omega = A\tilde{V}$.
	
In our problem, each failure has a covariance matrix $C_p$, where $p$ is the percentage of failure. We perform the PCA on $\bar{C} = \int C_p \delta p$, in order to obtain the PC independent of $p$.

\textbf{Most relevant principal components of $A$.} In order to choose the $l$ most relevant principal components, matrices $V$ (eigenvectors) and $D$ (eigenvalues) must be column-sorted in decreasing order, according to the eigenvalues in the diagonal of $D$. The importance of each eigenvector is characterized by its energy quantum $g$. The eigenvalues represent the distribution of the energy of $A$ among each of the eigenvectors. The energy quantum for the $j_{th}$ eigenvector is the sum of the energy quantum across all eigenvalues from 1 to $j$:
\begin{equation}
	g[j] = \sum_{k=i}^{j} D[k][k] \quad j=1..n \quad .
\end{equation}

Let $\tilde{V}$ be an $n \times l$, where $l \leq n$ matrix that contains the most relevant eigenvectors. Then, the objective is to choose an $l$ value as low as possible while preserving a reasonable high value of $g$ on a percentage basis. For instance, we have chosen $l$ so that $g$ is above a certain threshold $\alpha$:
\begin{equation}
	min\{l \in [1..n]: \frac{g[l]}{g[n]} \geq \alpha\}
\end{equation}
In this work we have considered $\alpha = 0.9$, from which we have obtained $l=1$.

\textbf{Simulation details.} The computation of each metric has been done with PHISON \cite{phison2012spects}. The simulations were performed on a Linux system with a 16-core 64-bit Intel Xeon processor of 2Ghz and 64 GB of RAM. The presented results are the average of 500 and 100 differently seeded simulation runs for random and targeted failures, respectively. The figures have been plotted by means of the \emph{pcolor} function of MATLAB. In addition, the PCA has also been done with MATLAB.

\section*{Acknowledgments} 
This work is partially supported by the Spanish Ministry of Science and Innovation project TEC 2012-32336, and by the Generalitat de Catalunya research support program SGR-1202. This work is also partially supported by the Secretariat for Universities and Research (SUR) and the Ministry of Economy and Knowledge through AGAUR FI-DGR 2012 and BE-DGR 2012 grants.

\bibliographystyle{naturemag}


\begin{table}[h]
%\small
\caption{Definition of the variables.}
\label{tab:var_def}
%\resizebox{\textwidth}{!}
\begin{center}
	\begin{tabular}{|c|p{9.7cm}|}
  \hline
  Variable & Meaning \\ \hline\noalign{\smallskip}\hline
 $n$ & number of robustness metrics\\
 $R$ & $R$-value \cite{pvannieghemframework}\\
 $s$ & vector of weights (size $n \times 1$)\\
 $t$ & vector of metrics (size $n \times 1$)\\
 $m$ & failure configurations, i.e., different realizations of the failure process\\
 $R^*$ & $R$-value computed via Principal Components (PC)\\
 $t^0$ & vector of metrics without failures (size $n \times 1$)\\
 $R^{*}_{init}$ & initial $R^*$-value (without failures)\\
 $v$ & eigenvector PC (size $n \times 1$)\\
 $\hat{v}$ & normalized eigenvector PC (size $n \times 1$)\\
 $P$ & set of percentage of failures\\
 $p$ & percentage of failures ($p \in P$)\\
 $t^p$ & vector of metrics when $p$\% of elements fail\\
 $R^{*}_p$ & $R$-value when $p$\% of elements fail\\
 $A_p$ & $m \times n$ matrix, i.e., $m$ values for each of the $n$ metrics when $p$\% of elements fail\\
 $\omega_p$ & vector of $R^{*}_p$ values (size $m \times 1$)\\
 ${\omega}'_p$ & vector $\omega_p$ sorted in decreasing order\\
 $C_p$ & covariance matrix of $A_p$ (size $n \times n$)\\
 $\bar{C}$ & average of the $|P|$ covariance matrices (size $n \times n$)\\
 $V$ & matrix containing $n$ eigenvectors $v$\\
 $D$ & diagonal matrix with eigenvalues (size $n \times n$)\\
 $l$ & number of most relevant eigenvector\\
 $\Omega$ & robustness surface, i.e., $|P|$ vectors ${\omega}'_p$\\
  \hline
\end{tabular}
\end{center}
\end{table}

\begin{table}
%\small
\caption{Main network characteristics. The table displays, from left to right, topology name, number of nodes ($N$), number of links ($L$), average node degree $\pm$
 \textit{standard deviation} (StDev) ($\langle k \rangle$), maximum degree ($k_{\max}$), average shortest-path length $\pm$ StDev ($\langle l \rangle$) and assortativity ($r$).\label{tab:features}}
%\resizebox{\textwidth}{!}
{\begin{center}\begin{tabular}{|l|c|c|c|c|c|c|}
  \hline
  \textit{topology} & $N$ & $L$ & $\langle k \rangle$ $\pm$ StDev & $k_{\max}$ & $\langle l \rangle$ $\pm$ StDev & $r$ \\ \hline\noalign{\smallskip}\hline
\emph{sprailway}    &   169 & 190     &   $2.24\; {\pm}1.09$   & 8 &  $10.49\; {\pm}4.64$ & -0.269 \\
\emph{europg}         &   1,494 & 2,154     &   $2.88\; {\pm}1.75$   & 13 &  $18.88\; {\pm}8.73$ & -0.119 \\  %\hline\noalign{\smallskip}\hline
%\emph{er199}&  199  &  597  &  $6\; {\pm}2.33$     &    16   &  $3.13\; {\pm}0.81$   &   0.0102     \\
%\emph{er400}&   400   &   2,367   &   $11.83\; {\pm}3.51$     &   27  &   $2.68\; {\pm}0.58$    & 0.044       \\
  \hline
\end{tabular}\end{center}}
\end{table}

\end{document}